\documentclass[pre,aps,superscriptaddress,floatfix,twocolumn]{revtex4-1}
\usepackage{amsmath,amssymb,multirow,epsfig,bm,pifont}
\usepackage{mathtools, float}
\usepackage[toc,page]{appendix}
\usepackage{comment}

\newcommand{\beq}{\begin{equation}}
\newcommand{\eeq}{\end{equation}}
\newcommand{\bea}{\begin{eqnarray*}}
\newcommand{\eea}{\end{eqnarray*}}

\usepackage[plainpages=false,pdfpagelabels,colorlinks=true,linkcolor=red,urlcolor=blue,citecolor=blue,pdftitle={},pdfauthor={},pdfdisplaydoctitle=true,pdfduplex=DuplexFlipLongEdge]{hyperref}

\begin{document} 

\title{Entanglement and matrix elements of observables in interacting integrable systems}
\author{Tyler LeBlond}
\affiliation{Department of Physics, The Pennsylvania State University, University Park, Pennsylvania 16802, USA}
\author{Krishnanand Mallayya}
\affiliation{Department of Physics, The Pennsylvania State University, University Park, Pennsylvania 16802, USA}
\author{Lev Vidmar}
\affiliation{Department of Theoretical Physics, J. Stefan Institute, SI-1000 Ljubljana, Slovenia}
\affiliation{Department of Physics, Faculty of Mathematics and Physics, University of Ljubljana, SI-1000 Ljubljana, Slovenia}
\author{Marcos Rigol}
\affiliation{Department of Physics, The Pennsylvania State University, University Park, Pennsylvania 16802, USA}

\begin{abstract}
We study the bipartite von Neumann entanglement entropy and matrix elements of local operators in the eigenstates of an interacting integrable Hamiltonian (the paradigmatic spin-1/2 XXZ chain), and we contrast their behavior with that of quantum chaotic systems. We find that the leading term of the average (over all eigenstates in the zero magnetization sector) eigenstate entanglement entropy has a volume-law coefficient that is smaller than the universal (maximal entanglement) one in quantum chaotic systems. This establishes the entanglement entropy as a powerful measure to distinguish integrable models from generic ones. Remarkably, our numerical results suggest that the volume-law coefficient of the average entanglement entropy of eigenstates of the spin-1/2 XXZ Hamiltonian is very close to, or the same as, the one for translationally invariant quadratic fermionic models. We also study matrix elements of local operators in the eigenstates of the spin-1/2 XXZ Hamiltonian at the center of the spectrum. For the diagonal matrix elements, we show evidence that the support does not vanish with increasing system size, while the average eigenstate-to-eigenstate fluctuations vanish in a power-law fashion. For the off-diagonal matrix elements, we show that they follow a distribution that is close to (but not quite) log-normal, and that their variance is a well-defined function of $\omega=E_{\alpha}-E_{\beta}$ ($\{E_{\alpha}\}$ are the eigenenergies) proportional to $1/D$, where $D$ is the Hilbert space dimension.
\end{abstract}
\date{\today}

\maketitle
\section{Introduction}

Much work has been done in the past decade to understand the far-from-equilibrium dynamics and description after equilibration of isolated nonintegrable (generic) and integrable quantum many-body systems~\cite{ETH_review, eisert_friesdorf_review_15, polkovnikov_sengupta_review_11}. Despite tremendous progress in recent years~\cite{essler_fagotti_2016, calabrese_cardy_2016, cazalilla_chung_2016, bernard_doyon_2016, caux_2016, GGE_review, ilievski_medenjak_2016, langen_gasenzer_2016, vasseur_moore_2016, deluca_mussardo_2016}, the microscopics of interacting integrable systems are those that remain less understood. On the one hand, interactions are present in those systems as in nonintegrable ones (making their study challenging), and on the other hand, they exhibit extensive numbers of local conserved quantities as noninteracting systems do. The presence of such quantities precludes thermalization in integrable (noninteracting and interacting) quantum systems~\cite{Kinoshita2006, GGE_Rigol_2007, wouters_denardis_14, pozsgay_mestyan14, ilievski_denardis_15}.

Thermalization does occur in generic isolated quantum systems, and is understood to be a consequence of quantum chaos and eigenstate thermalization~\cite{Deutsch1991, Srednicki1994, ETHRigol, ETH_review}. Essentially, the matrix elements of observables (few-body operators) $\hat{O}$ in eigenstates of generic quantum Hamiltonians are described by the eigenstate thermalization hypothesis (ETH) ansatz~\cite{ETH_review,Srednicki_1999}:
\beq \label{eq:ETH} 
O_{\alpha\beta}=O(\bar{E})\delta_{\alpha\beta}+e^{-S(\bar{E})/2}f_O(\bar{E},\omega)R_{\alpha\beta},
\eeq
where $\bar{E}\equiv (E_{\alpha}+E_{\beta})/2$, $\omega=E_{\alpha}-E_{\beta}$, and $S(\bar E)$ is the thermodynamic entropy at energy $\bar E$. The functions $\mathcal{O}(\bar{E})$ and $f_O(\bar{E},\omega)$ are smooth, and $R_{\alpha\beta}$ is a normally distributed variable with zero mean and unit variance (variance 2) for $\alpha \neq \beta$ ($\alpha=\beta$) in Hamiltonians that exhibit time-reversal symmetry. The smoothness of the diagonal matrix elements allows observables described by Eq.~(\ref{eq:ETH}) to thermalize (i.e., to be described by traditional ensembles of statistical mechanics) for experimentally relevant initial conditions. The off-diagonal matrix elements control the approach to and fluctuations about equilibrium~\cite{ETH_review}. 

The ETH ansatz~(\ref{eq:ETH}) has been extensively tested in exact diagonalization studies of nonintegrable Hamiltonians~\cite{ETH_review, ETHRigol, Breakdown, rigol_09b, Santos2010Localization, Khatami, Steinigeweg2013Eigenstate, Beug1, steinigeweg_khodja_14, sorg_vidmar_14, Kim2014Testing, Beug2, 2DTFIM, 2DTFIM-2, Sagawa2018, khaymovich_haque_19, Jansen2019Eigenstate, mierzejewski_vidmar_19}. While the diagonal matrix elements display a shrinking support and an exponentially decaying variance with increasing system size in nonintegrable systems, confirming that the diagonal ETH is valid for them, the same is not true in integrable systems~\cite{Breakdown, rigol_09b, Santos2010Localization, Beug1, GGE_Rigol_2011, GGE_review, mierzejewski_vidmar_19}. In integrable systems, the support of the diagonal matrix elements need not shrink, and their variance is expected to decay as a power law in system size~\cite{PhysRevLett.105.250401, Ikeda, Alba}. The latter is consistent with the fact that the variance of the diagonal matrix elements of intensive translationally invariant observables must vanish at least as a power law in system size because it is bounded from above by the Hilbert-Schmidt norm (which vanishes as a power law in system size)~\cite{mierzejewski_vidmar_19}.
 
The off-diagonal matrix elements of observables in the eigenstates of interacting integrable quantum many-body systems have received little attention. While such results have been reported for specific models and system sizes alongside those of quantum chaotic systems~\cite{rigol_09b, Santos2010Localization, Beug2}, there has been no systematic study of their properties. For noninteracting models (or models mappable to them), the existence of an increasingly large (with increasing system size) fraction of vanishing off-diagonal matrix elements precludes the definition of a meaningful function $f_O(\bar{E},\omega)$, in contrast to quantum chaotic models~\cite{Khatami}. On the other hand, recent results by two of us (K.M. and M.R.) in the context of periodically driven systems provided strong evidence that one can define (and experimentally measure) a function $|f_O(\bar{E},\omega)|^2=e^{S(\bar{E})}\overline{|O_{\alpha\beta}|^2}$ for interacting integrable models~\cite{1907.04261}. Exploring this, along with other properties of the off-diagonal (and diagonal) matrix elements in the spin-1/2 XXZ chain, is one of the two central goals of this work. 

The other central goal of this work is to study the structure of highly excited energy eigenstates by means of their bipartite entanglement. Recently, much work has been devoted to understanding entanglement properties of highly-excited eigenstates of many-body Hamiltonians~\cite{alba09, deutsch_10, santos_polkovnikov_12, hamma_santra_12, deutsch13, moelter_barthel_14, storms14, beugeling15, yang_chamon_15, lai15, nandy_sen_16, ee_fermion, ee_chaotic, dymarsky_lashkari_18, riddell_muller_18, zhang_vidmar_18, liu_chen_18, garrison_grover_18, nakagawa_watanabe_18, vidmar_hackl_18, huang_19, hackl_vidmar_19, murciano_ruggiero_19, lu_grover_19, sun_nie_19, giovenale_pont_19,  bertini_kos_19, huang_gu_19, murthy_srednicki_19, morampudi_chandran_18, modak_nag_19, miao_barthel_19, bianchi_dona_19, faiez_sefranek_19, faiez_sefranek_19b, jafarizadeh_rajabpour_19}. Here we study the average entanglement entropy over all eigenstates of the spin-1/2 XXZ Hamiltonian in the zero-magnetization sector. We argue that this average universally reveals the fundamentally different natures of interacting integrable and quantum chaotic models. While in both nonintegrable and interacting integrable systems the leading term of the average entanglement entropy exhibits a volume-law scaling, we show that the corresponding volume-law coefficient is markedly different between the two. In quantum chaotic systems~\cite{ee_chaotic} it matches the prediction by Page~\cite{page} for random pure states in the Hilbert space, while it is smaller for interacting integrable systems. Remarkably, our results for interacting spin-1/2 integrable systems are consistent with this coefficient being very close to, or the same as, the one for translationally invariant free~\cite{ee_fermion} and (more generally) quadratic~\cite{vidmar_hackl_18, hackl_vidmar_19} fermionic Hamiltonians. This suggests that, entanglement-wise, the overwhelming majority of the eigenstates are very similar between interacting spin-1/2 integrable Hamiltonians and noninteracting fermionic Hamiltonians.

The presentation is organized as follows: In Sec.~\ref{model}, we discuss the specific integrable and nonintegrable models and observables considered, as well as details about the numerical calculations carried out. In Sec.~\ref{entang}, we compare the average entanglement entropy of eigenstates of the spin-1/2 XXZ chain with that of eigenstates of noninteracting and nonintegrable models. In Sec.~\ref{diag}, we discuss the distributions and scaling properties of the diagonal matrix elements of two local observables at the center of the spectrum. In Sec.~\ref{offdiag}, we discuss the off-diagonal matrix elements of the same observables: their distributions, scaling properties, and functional dependence of $|f_O(\bar{E},\omega)|^2$ on $\omega$, for $\bar{E}$ at the center of the spectrum. Lastly, in Sec.~\ref{conc}, we summarize our results.

\section{Model}\label{model}

We study the spin-1/2 XXZ chain with the addition of next-nearest-neighbor interactions, with $L$ sites and periodic boundary conditions. The Hamiltonian is
\begin{eqnarray}\label{eq:xxz} 
\hat{H} &=& \sum_{i=1}^L \left[ \frac{1}{2} \left( \hat{S}_i^+\hat{S}_{i+1}^- + \text{H.c.} \right) + \Delta \hat{S}_i^z\hat{S}_{i+1}^z \right]\nonumber \\
&&+ \lambda \sum_{i=1}^L \left[ \frac{1}{2} \left( \hat{S}_i^+\hat{S}_{i+2}^- + \text{H.c.} \right) + \frac{1}{2} \hat{S}_i^z\hat{S}_{i+2}^z \right] ,
\end{eqnarray}
where $\hat{S}_i^{\nu}$ are spin-1/2 operators in the $\nu\in\{x,y,z\}$ directions on site $i$, and $\hat{S}_i^{\pm} = \hat{S}_i^x\pm i\hat{S}_i^y$ are the corresponding ladder operators. When $\lambda=0$,  Hamiltonian~\eqref{eq:xxz} is integrable and can be solved exactly using the Bethe ansatz \cite{RevModPhys.83.1405}. When $\lambda\ne0$, Hamiltonian~\eqref{eq:xxz} is quantum chaotic~\cite{PhysRevE.81.036206}. We set $\lambda=0$ and $1$ to compare the integrable and nonintegrable regimes, respectively. Unless otherwise specified, we show results for $\Delta=0.55$ and $\Delta=1.1$ to illustrate that they are qualitatively similar in the (nearest-neighbor) easy-plane ($\Delta<1$) and easy-axis ($\Delta>1$) regimes. 

We study the matrix elements of two local operators: The nearest-neighbor $z$-interaction
\beq \label{eq:obs_A}
\hat{A}=\dfrac{1}{L}\sum_{i=1}^L \hat{S}^z_{i}\hat{S}^z_{i+1},
\eeq
and the next-nearest-neighbor flip-flop operator 
\beq \label{eq:obs_B}
\hat{B}=\dfrac{1}{L}\sum_{i=1}^L \left(\hat{S}^+_i\hat{S}^-_{i+2}+\text{H.c.}\right).
\eeq

To study the entanglement entropy of energy eigenstates, as well as the matrix elements of $\hat{A}$ and $\hat{B}$ in those eigenstates, it is important to resolve all the symmetries of the Hamiltonian~\cite{ETH_review, Santos2010Localization}. First, we note that Hamiltonian~\eqref{eq:xxz} conserves the total magnetization in the $z$-direction, $M^z = \sum_i\hat{S}_i^z$. In this work, we focus on the zero magnetization sector in chains with even numbers of lattice sites. The next important symmetry is translation, which allows one to block-diagonalize the Hamiltonian in different total quasimomentum $k$ sectors. All quasimomentum sectors are used in the average entanglement entropy calculations reported in Sec.~\ref{entang}. Within the $M^z=0$ and $k=0$ sector, there are two additional symmetries, namely, spin inversion ($Z_2$) and space reflection ($P$). In our studies of matrix elements, we focus on the even-$Z_2$ even-$P$ sector within the $M^z=0$ and $k=0$ sector. We use full exact diagonalization of periodic chains with up to $L=26$ sites. The even-$Z_2$ even-$P$ sector within the $M^z=0$ and $k=0$ sector of the chain with $L=26$, the largest considered, has $101,340$ states.

\section{Entanglement Entropy} \label{entang}

In this section, we study the entanglement properties of eigenstates $\{ |\alpha\rangle \}$ of Hamiltonian~(\ref{eq:xxz}) in the zero magnetization sector. We consider a bipartition into a subsystem $A$ and its complement $\bar A$ that consist of $L_{\rm A}$ and $L-L_{\rm A}$ consecutive lattice sites, respectively. We calculate the bipartite entanglement entropy of an eigenstate $|\alpha\rangle$ as
\begin{equation}
 S_\alpha = - {\rm Tr}\{ \hat \rho_A \ln(\hat \rho_A) \} \, ,
\end{equation}
where $\hat \rho_A = {\rm Tr}_{\bar A}\{ |\alpha\rangle \langle \alpha| \}$ is the reduced density matrix of the subsystem $A$. We average $S_\alpha$ over all Hamiltonian eigenstates in the zero magnetization sector to obtain the average entanglement entropy $\bar S = {\cal D}^{-1} \sum_\alpha S_\alpha$, where ${\cal D} = \binom{L}{L/2}$.

\begin{figure}[!]
\centering \includegraphics[width=1.0\columnwidth]{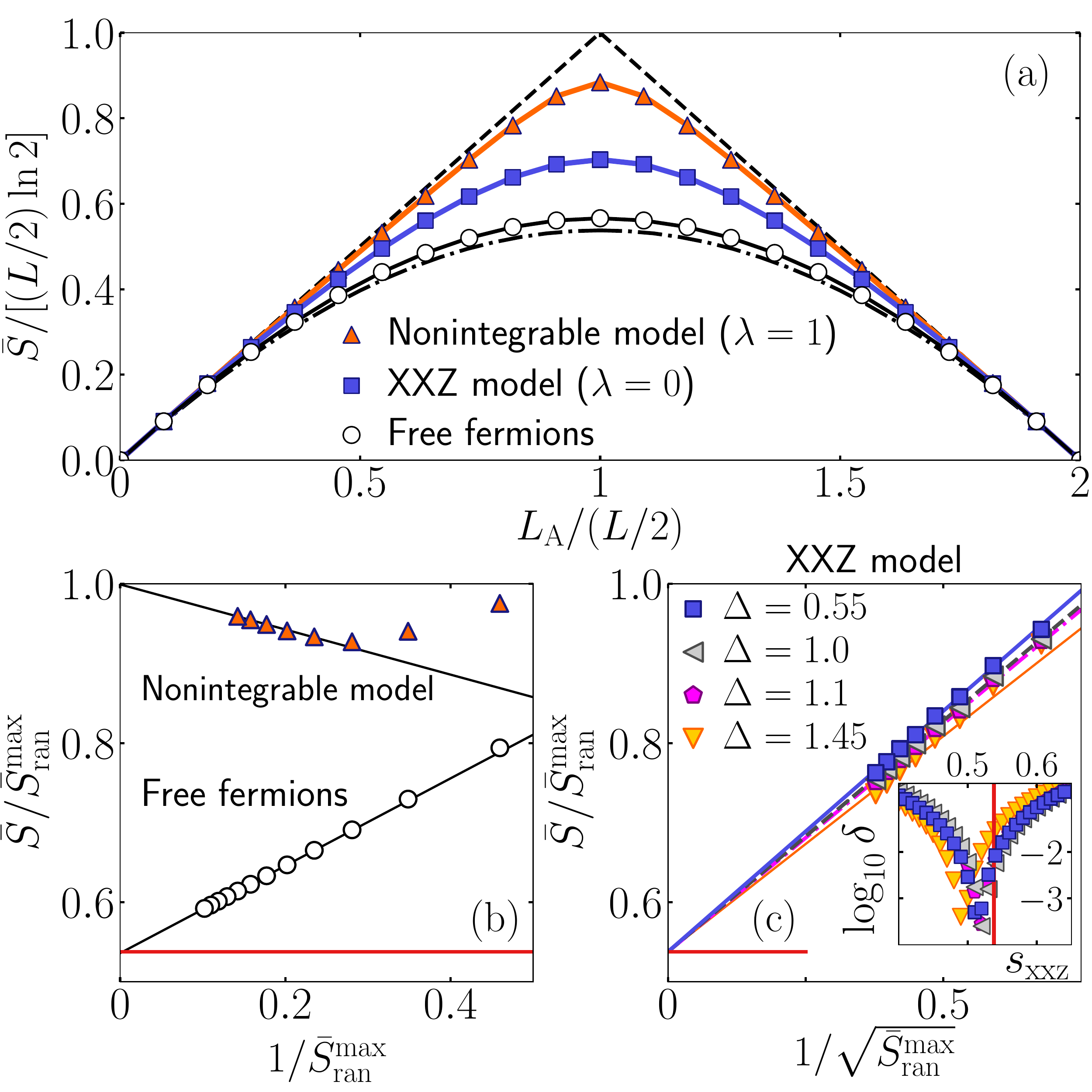}
\vspace{-0.5cm}
\caption{\label{fig:entanglement} Average entanglement entropy $\bar S$, in the zero-magnetization sector (half-filling for free fermions), for three paradigms of many-body quantum systems: the spin-1/2 XXZ model [$\lambda = 0$ in Eq.~(\ref{eq:xxz}), an interacting integrable model], a nonintegrable model [$\lambda = 1$ in Eq.~(\ref{eq:xxz})], and free fermions [which are mappable to the spin-1/2 XX spin chain, Eq.~(\ref{eq:xxz}), with $\Delta = \lambda = 0$]. (a) $\bar S$ vs $L_{\rm A}$ for $L=22$ ($\Delta=0.55$ in the interacting models). The dashed line shows the results for random pure states from Eq.~\eqref{def_Smax} in the thermodynamic limit. The dashed-dotted line is the average for free fermions in all particle sectors at $L=36$ (the same results reported in Fig.~1 in Ref.~\cite{ee_fermion}). Panels (b) and (c) show finite-size scaling analyses at $L_{\rm A}/L = 1/2$. We normalize $\bar S$ by $\bar S_{\rm ran}^{\rm max} \equiv \bar S_{\rm ran}\left(\frac L2,\frac{1}{2}\right)$, see Eq.~(\ref{def_Smax}), and plot the results vs $\bar S_{\rm ran}^{\rm max}$ to show that $\bar S = \bar S_{\rm ran}^{\rm max}$ for the nonintegrable model in the thermodynamic limit and to improve the scaling analyses in our small systems. (b) Normalized averages as functions of $1/\bar S_{\rm ran}^{\rm max}$ at the nonintegrable point and for free fermions. Lines are two-parameter fits to $c_0+c_1/\bar S_{\rm ran}^{\rm max}$ with $c_0=0.999$ (with 1 minus the coefficient of determination, $1-R^2=4.3\times 10^{-3}$) and $c_0+c_1/\bar S_{\rm ran}^{\rm max}$ with $c_0=0.536$ (with $1-R^2=2.5\times 10^{-5}$), respectively, for $L\geq 14$. (c) Normalized averages at $\Delta = 0.55$, 1.0, 1.1, and 1.45 in the spin-1/2 XXZ model as functions of $1/\sqrt{\bar S_{\rm ran}^{\rm max}}$. Lines are single-parameter fits to the function $s_{\rm free}\left( \frac{1}{2}\right) + d_1/\sqrt{\bar S_{\rm ran}^{\rm max}}$ for systems with $L \geq 14$. Inset in (c), $\delta=1-R^2$ of the fits to $s_{\rm xxz} + d_1/\sqrt{\bar S_{\rm ran}^{\rm max}}$ as a function of the volume-law coefficient $s_{\rm xxz}$ chosen. Note that the minima are very close to $s_{\rm free}\left( \frac{1}{2}\right)$ for $\Delta\simeq 1$, and depart from that value as the quality of the fit worsens when departing from $\Delta=1$ (presumably because of stronger finite-size effects).} 
\end{figure}

The upper bound for the entanglement entropy of pure states in a given magnetization sector (or, equivalently, in a given particle-number sector when mapping spin-1/2 systems onto hard-core bosons or spinless fermions), for a given $L_{\rm A}/L$, depends both on the magnetization and on $L_{\rm A}/L$. The leading term, which scales with the volume, depends on the magnetization~\cite{ee_chaotic, garrison_grover_18}. There is also a subleading, $O(1)$, term that depends on $L_{\rm A}/L$~\cite{ee_chaotic}. 

In the zero magnetization sector, for $L_A\leq L/2$, the leading and first subleading terms in the average entanglement entropy of random pure states with normally distributed real coefficients are~\cite{ee_chaotic}
\begin{eqnarray} \label{def_Smax}
&&\bar S_{\rm ran}\left(L_{\rm A},\frac{L_{\rm A}}{L}\right)\nonumber \\ && = L_{\rm A} \ln(2) + \frac{\frac{L_{\rm A}}{L} + \ln(1-\frac{L_{\rm A}}{L})}{2} - \frac{1}{2} \frac{\sum_{n_{\rm A}=0}^{L_{\rm A}} \binom{L_{\rm A}}{n_{\rm A}}^2}{\binom{L}{L/2}} \, \nonumber \\
&& = L_{\rm A} \ln(2) + \frac{\frac{L_{\rm A}}{L} + \ln(1-\frac{L_{\rm A}}{L})}{2} - \frac{1}{2} \frac{\binom{2L_{\rm A}}{L_{\rm A}}}{\binom{L}{L/2}} \, .
\end{eqnarray}
On the r.h.s.~of Eq.~(\ref{def_Smax}), the first two terms are the upper bound for the entanglement entropy of pure states in the $M^z = 0$ sector~\cite{ee_chaotic}, while the third term is the generalization of the correction derived by Page~\cite{page} for the $M^z = 0$ sector of a system with conserved $M^z$. Motivated by the numerical results in Ref. \cite{ee_chaotic}, we think of $\bar S_{\rm ran}\left(L_{\rm A},\frac{L_{\rm A}}{L}\right)$ as an upper bound for the average entanglement entropy over all eigenstates of any given physical Hamiltonian, with $\bar S_{\rm ran}^{\rm max} \equiv \bar S_{\rm ran}\left(\frac L2,\frac{1}{2}\right)$ as the maximum. The dashed line in Fig.~\ref{fig:entanglement}(a) shows $\bar S_{\rm ran}\left(L_{\rm A},\frac{L_{\rm A}}{L}\right)$ in the thermodynamic limit.

On the opposite (low-entropy) side of physical Hamiltonians, one has noninteracting (free) fermions. Translationally invariant free fermionic Hamiltonians exhibit the same leading term of the average entanglement entropy as the XX model, Eq.~(\ref{eq:xxz}), with $\Delta = \lambda = 0$~\cite{hackl_vidmar_19}. It was proved in Ref.~\cite{ee_fermion} that the leading (volume) term of the average entanglement entropy over all (i.e., including all particle-number sectors) eigenstates in those models is
\begin{equation} \label{def_Sfree}
\bar S_{\rm free}\left(L_{\rm A},\frac{L_{\rm A}}{L}\right) = s_{\rm free} \left( \frac{L_{\rm A}}{L}\right) \, L_{\rm A} \ln 2 ,
\end{equation}
with $s_{\rm free}\left(0\right)=1$, and $0<s_{\rm free} \left( \frac{L_{\rm A}}{L}\right)<1$ for $L_{\rm A}/L > 0$. In Ref.~\cite{ee_fermion}, $s_{\rm free} \left( \frac{L_{\rm A}}{L}\right)$ was computed numerically [dashed-dotted line in Fig.~\ref{fig:entanglement}(a)] and, for $L_{\rm A}=L/2$, it was found that $s_{\rm free} \left( \frac{1}{2}\right)= 0.5378(1)$. Subsequently, it was conjectured that $s_{\rm free} \left( \frac{L_{\rm A}}{L}\right)$ is universal for all translationally invariant quadratic fermionic models~\cite{vidmar_hackl_18, hackl_vidmar_19}. The horizontal lines in Figs.~\ref{fig:entanglement}(b) and~\ref{fig:entanglement}(c) show $s_{\rm free} \left( \frac{1}{2}\right)/\bar S_{\rm ran}^{\rm max}$. 

\begin{figure*}[!t]
\centering \includegraphics[width=0.95\textwidth]{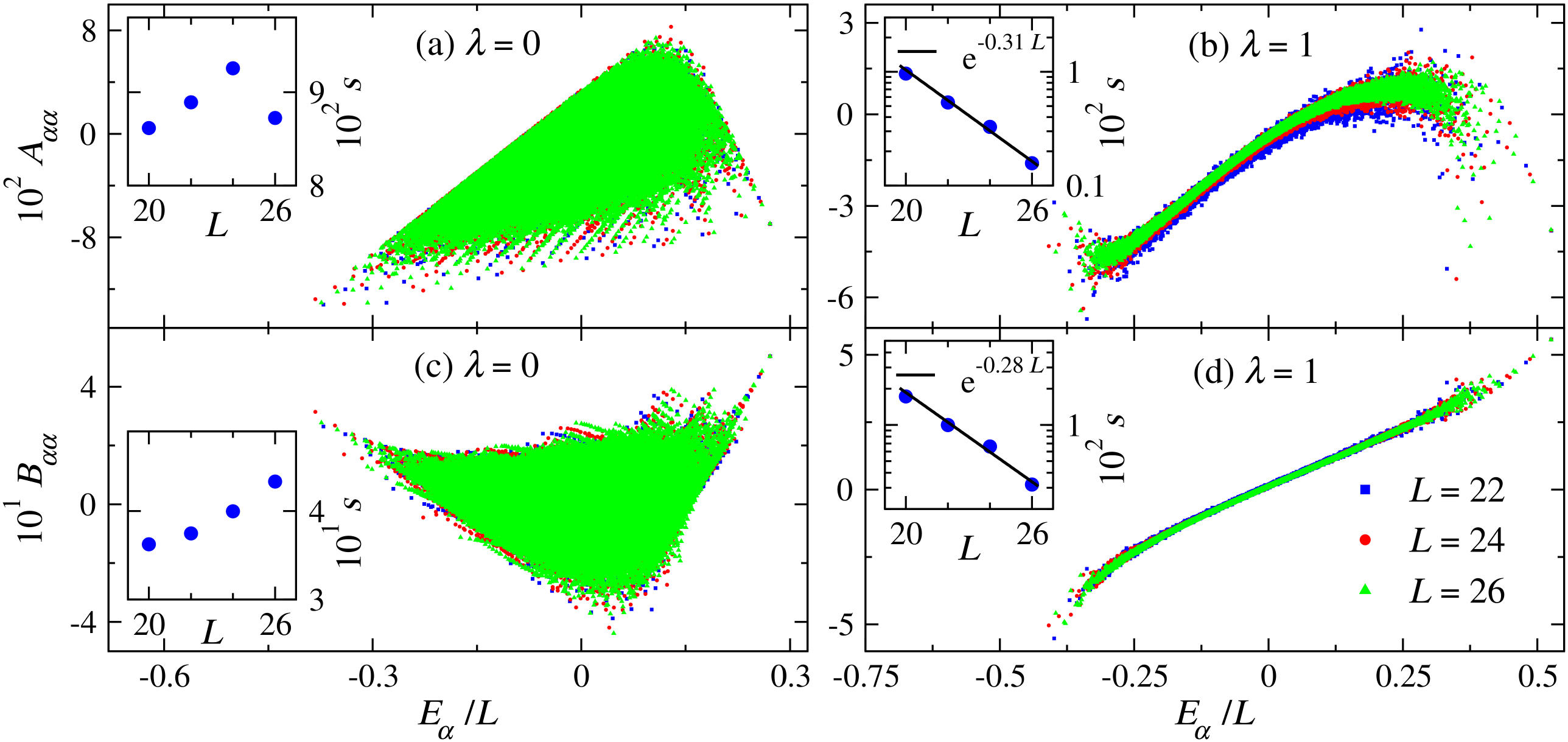}
\caption{\label{fig:eev_overlap} EEVs ($O_{\alpha\alpha}$) of local observables $\hat{A}$ [(a) and (b), see Eq.~\eqref{eq:obs_A}] and $\hat{B}$ [(c) and (d), see Eq.~\eqref{eq:obs_B}] at integrable [(a) and (c), $\lambda=0$] and nonintegrable [(b) and (d), $\lambda=1$] points of Hamiltonian~\eqref{eq:xxz}, plotted vs the eigenstate energies per site $E_{\alpha}/L$, for each eigenstate $|\alpha\rangle$. Results are shown for chains with $L=22$, 24, and $26$ sites, and are superposed to demonstrate non-shrinking (shrinking) support in the integrable (nonintegrable) case. $\Delta=0.55$ in all plots (qualitatively similar results were obtained, not shown, for other values of $\Delta$). (Insets) A quantifier for the support of the EEVs, $s=\max(\left|O_{\alpha\alpha} - \overline{O_{\alpha\alpha}}\right|)$, where $\overline{O_{\alpha\alpha}}$ is the running average over 201 EEVs centered at $O_{\alpha\alpha}$ and $\max(\cdot)$ is taken over the central 50\% of the spectrum, plotted as a function of $L$.} 
\end{figure*}

In Fig.~\ref{fig:entanglement}(a), we show the average entanglement entropy over all eigenstates within the half-filled sector of noninteracting fermions, as well as within the zero-magnetization sector of integrable ($\Delta=0.55$ and $\lambda = 0$) and nonintegrable ($\Delta=0.55$ and $\lambda = 1$) points of Eq.~(\ref{eq:xxz}), for chains with $L=22$. The results at the nonintegrable point are closest to the thermodynamic limit ones for random pure states. Figure~\ref{fig:entanglement}(b), for $L_{\rm A}/L=1/2$, shows that the small differences seen in Fig.~\ref{fig:entanglement}(a) appear to vanish in the thermodynamic limit, in agreement with complementary results reported in Ref.~\cite{ee_chaotic}. For free fermions, on the other hand, the results in Fig.~\ref{fig:entanglement}(a) are closest to the thermodynamic limit ones obtained in Ref.~\cite{ee_fermion} by averaging over all fillings. Figure~\ref{fig:entanglement}(b), for $L_{\rm A}/L=1/2$, shows that the differences seen in Fig.~\ref{fig:entanglement}(a) appear to vanish in the thermodynamic limit as expected (the zero magnetization sector is the one dominant in the thermodynamic limit when averaging over all fillings).

The numerical results at the interacting integrable point ($\Delta=0.55$) in Fig.~\ref{fig:entanglement}(a) are in between the ones for random pure states and the noninteracting ones. However, the finite-size scaling analysis reported in Fig.~\ref{fig:entanglement}(c) for $L_{\rm A}/L=1/2$ suggests that the leading term of the average entanglement entropy at $\Delta=0.55$ (easy-plane regime), is very close to, or the same as, the one for noninteracting fermions. As a matter of fact, finite-size scaling analyses in Fig.~\ref{fig:entanglement}(c) for $\Delta=1.1$, 1.45 (easy-axis regime) and $\Delta=1.0$ (Heisenberg point, the most symmetric point in the spin-1/2 XXZ model) suggest that this is true independently of the value of $\Delta$.

The finite-size scaling analyses in Figs.~\ref{fig:entanglement}(b) and~\ref{fig:entanglement}(c) suggest that the qualitatively new effect of interactions in integrable systems is subleading, as they change the first subleading term from $O(1)$ in noninteracting models [the leading correction to $\bar S /\bar S_{\rm ran}^{\rm max}$ in Fig.~\ref{fig:entanglement}(b) is $\propto 1/L_{\rm A}$] to $O(\sqrt{L_{\rm A}})$ in interacting integrable models [the leading correction to $\bar S /\bar S_{\rm ran}^{\rm max}$ in Fig.~\ref{fig:entanglement}(c) is $\propto 1/\sqrt{L_{\rm A}}$].

\section{Diagonal Matrix Elements} \label{diag}

In this section, we study expectation values of observables in eigenstates of interacting integrable and nonintegrable Hamiltonians, referred to in what follows as the eigenstate expectation values (EEVs) of the observables, in the even-$Z_2$ even-$P$ sector within the $M^z=0$ and $k=0$ sector (see Sec.~\ref{model}).

In Fig.~\ref{fig:eev_overlap}, we show the EEVs of observables $\hat{A}$ and $\hat{B}$ as functions of the eigenenergies per site ($E_{\alpha}/L$) for different chain sizes at integrable [Figs.~\ref{fig:eev_overlap}(a) and~\ref{fig:eev_overlap}(c)] and nonintegrable [Figs.~\ref{fig:eev_overlap}(b) and~\ref{fig:eev_overlap}(d)] points of Hamiltonian~\eqref{eq:xxz}. At the nonintegrable point, for both observables, one can see in the plots that the support (maximum spread) of the EEVs around each $E_{\alpha}/L$ (away from the edges of the spectrum) shrinks upon increasing the chain size $L$. The scaling of a quantifier of the support, see the insets in Figs.~\ref{fig:eev_overlap}(b) and~\ref{fig:eev_overlap}(d), indicates that the support vanishes exponentially fast with increasing $L$. This suggests that, in the thermodynamic limit, the EEVs are described by the smooth function $O(E)$, which, in turn, is the thermal expectation value of observable $\hat O$ at energy $E$~\cite{ETH_review}. Hence, one expects all EEVs at the nonintegrable point away from the edges of the spectrum to be thermal in the thermodynamic limit. On the other hand, at integrability, Figs.~\ref{fig:eev_overlap}(a) and~\ref{fig:eev_overlap}(c) show that the support of the EEVs is wide and does not shrink with increasing system size [see the insets in Figs.~\ref{fig:eev_overlap}(a) and~\ref{fig:eev_overlap}(c)]. The wide nonshrinking support indicates that at the integrable point ETH is not satisfied, as nonthermal states persist in the thermodynamic limit. The next question to address at the integrable point is how those EEVs are distributed.

\begin{figure}[!t]
\centering \includegraphics[width=1.0\columnwidth]{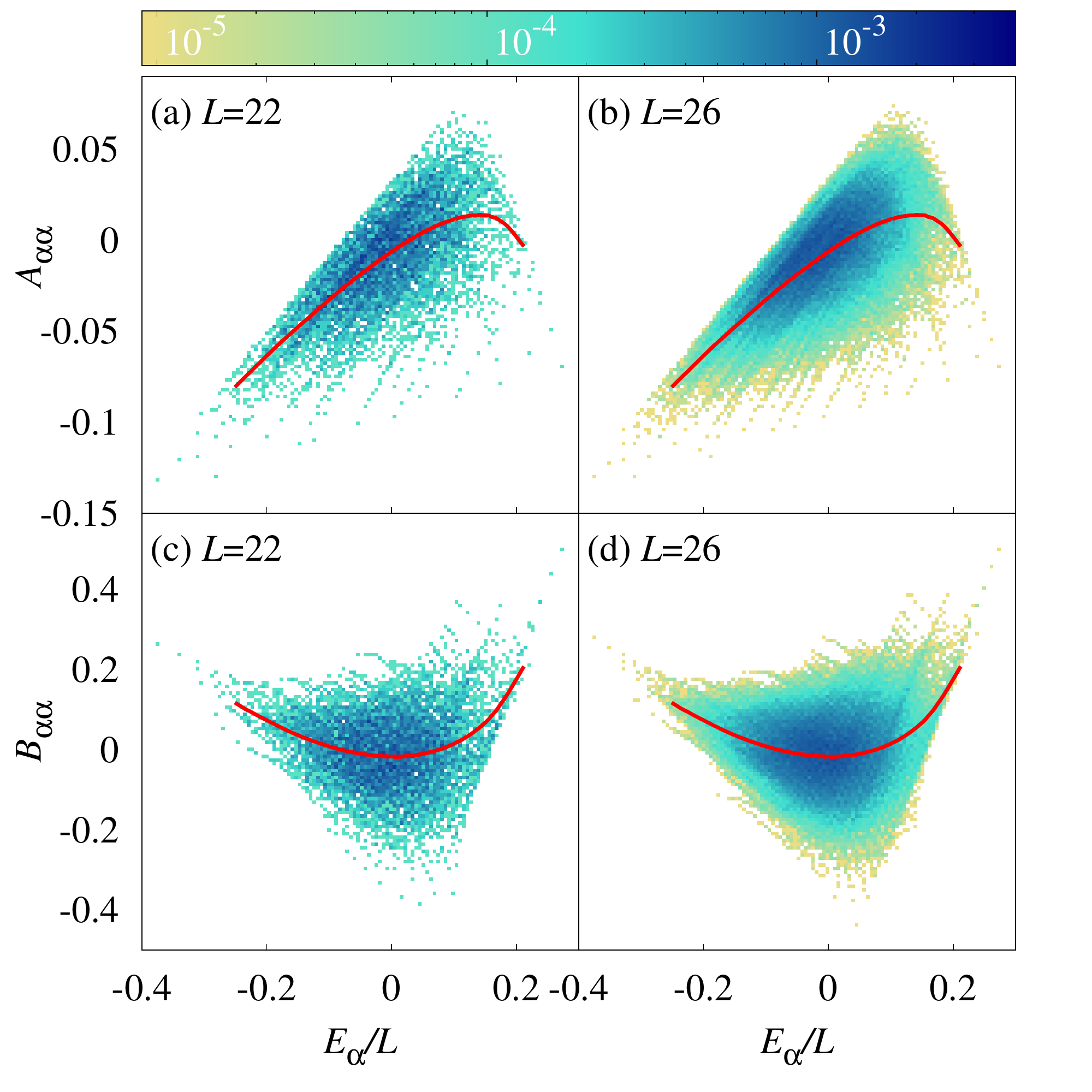}
\vspace{-0.6cm}
\caption{\label{fig:eev_heatmap} Normalized 2D histograms of the EEVs of $\hat{A}$ [(a) and (b)] and $\hat{B}$ [(c) and (d)] as functions of $E_{\alpha}/L$ at the integrable point ($\Delta=0.55$ and $\lambda=0$) of Hamiltonian (\ref{eq:xxz}). We show the microcanonical average (calculated for $L=26$ using $\delta E/L=0.05$) as a solid (red) line. Results are reported for $L=22$ [(a) and (c)] and $L=26$ [(b) and (d)].}
\end{figure} 

\begin{figure}[!t]
\centering \includegraphics[width=1.0\columnwidth]{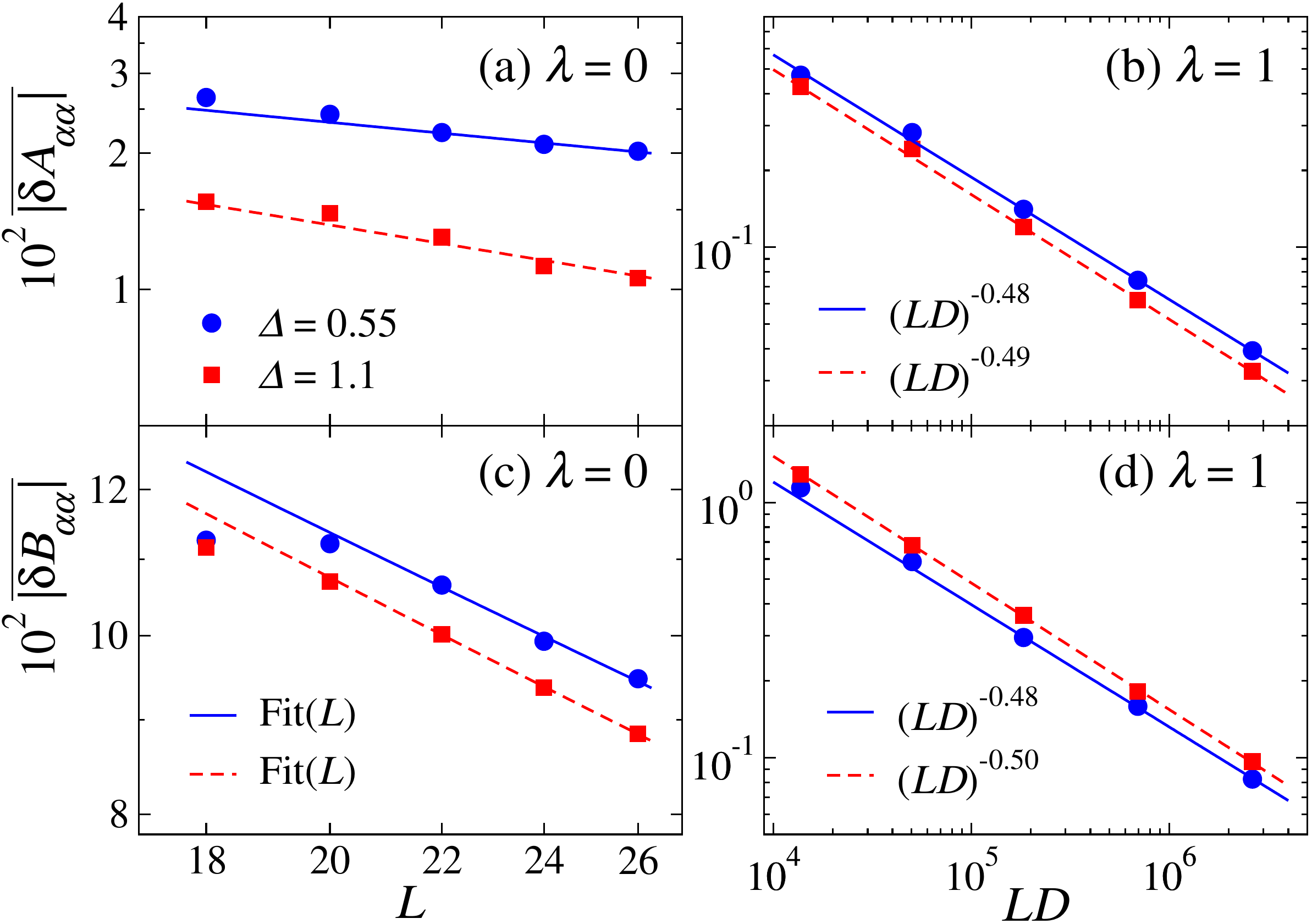}
\vspace{-0.5cm}
\caption{\label{fig:fluct_trend} Scaling of the average eigenstate-to-eigenstate fluctuations $\overline{|\delta O_{\alpha\alpha}|}$ [see Eq.~\eqref{eq:eev_fluct}] for $\hat{A}$ [(a) and (b)] and $\hat{B}$ [(c) and (d)] at two integrable [(a) and (c), $\lambda=0$] and two nonintegrable [(b) and (d), $\lambda=1$] points of Hamiltonian (\ref{eq:xxz}). We report results for $\Delta=0.55$ and $\Delta=1.1$. The symbols show the numerical results, while the lines depict fits to the functions $\text{fit}(L)=c_1/\sqrt{L}+c_2/L$ [(a) and (c)] and $c_1(LD)^{-c_2}$ [(b) and (d)] for $L=22$ through $L=26$.} 
\end{figure}

In Fig.~\ref{fig:eev_heatmap}, we show the normalized distribution (color coded) of the EEVs for observables $\hat{A}$ and $\hat{B}$ for two different system sizes ($L=22$ and 26) at the integrable point in Fig.~\ref{fig:eev_overlap}, along with the microcanonical averages (solid lines) for the respective observables. The microcanonical averages are calculated using the results from $L=26$ in an energy window $\delta E$ that is small enough to yield a smooth curve independent of $\delta E$. By comparing the results for $L=22$ and $L=26$ for each observable, one can see that, despite the large non-vanishing support, the distribution of EEVs (on the $y$-axis) becomes increasingly peaked (with increasing system size) about the microcanonical average (further evidence for this is reported in Fig.~\ref{fig:fluct_trend}). Similarly, on the $x$-axis, the distribution becomes increasingly peaked about the center of the spectrum ($E_\alpha/L=0$). The latter occurs because of the known Gaussian behavior of the density of states in local Hamiltonians~\cite{ETH_review, Hartmann2004, Hartmann2005}.

\begin{figure}[!t]
\centering \includegraphics[width=1.0\columnwidth]{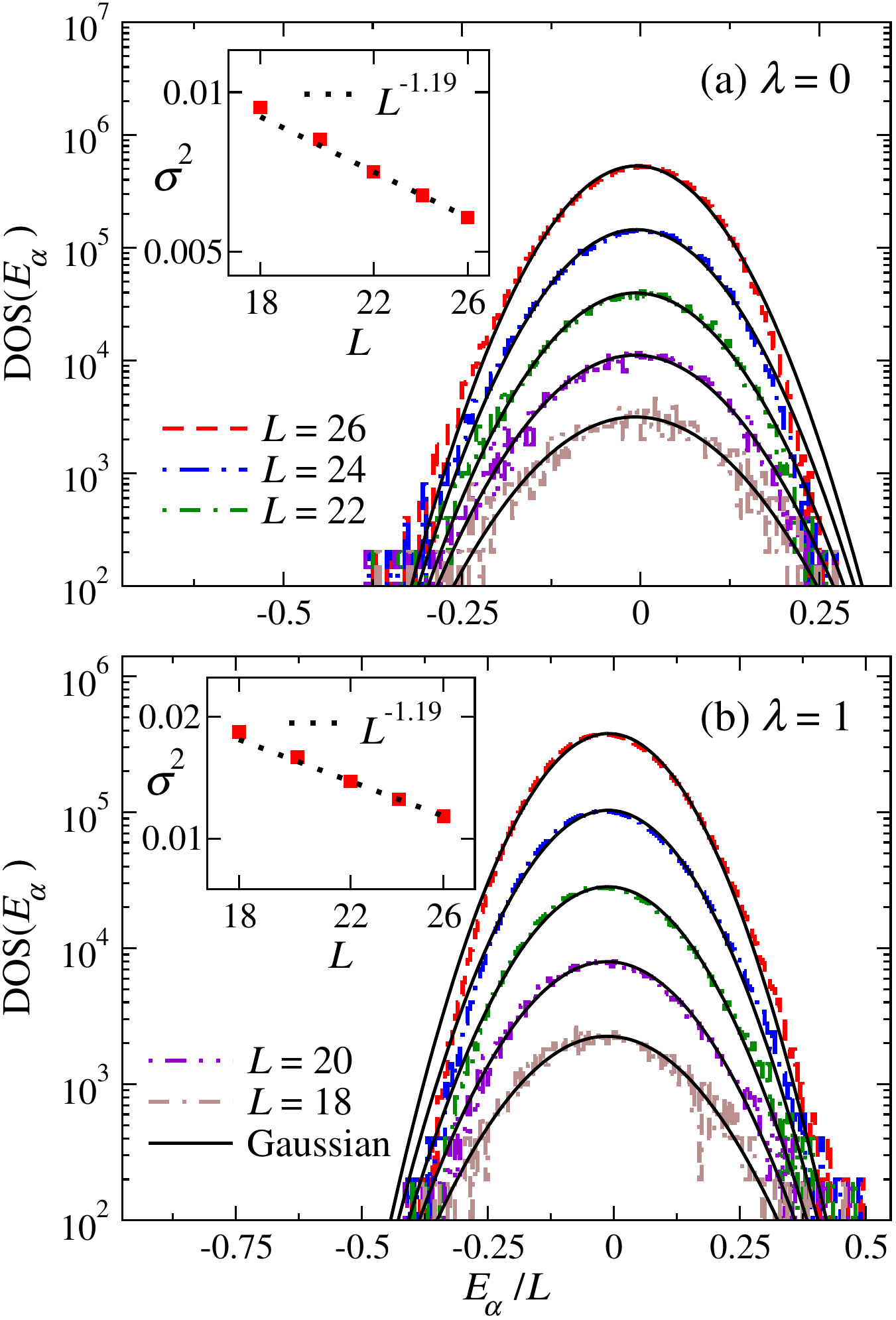}
\vspace{-0.5cm}
\caption{\label{fig:dos_fig} Densities of states (DOS) as functions of $E_{\alpha}/L$ for $L=18$ through $L=26$ at integrable [(a), $\lambda=0$] and nonintegrable [(b), $\lambda=1$] points ($\Delta=0.55$). The solid (black) lines are Gaussian functions with the same mean and variance as the data. (Insets) Variances of the data vs $L$, along with power law fits of the variances to $c_1 L^{-c_2}$.}
\end{figure} 

In Fig.~\ref{fig:dos_fig}, we plot the many-body density of states DOS($E_{\alpha}$) as a function of the energy per site $E_\alpha/L$ at integrable and nonintegrable points along with Gaussian functions that have  the same mean and variance as the data. The Gaussian functions agree well with the numerical results. This makes it apparent that, even for the small chains that one can solve using full exact diagonalization, the density of states is very close to a Gaussian function (away from the edges of the spectrum). This is true regardless of whether the Hamiltonian is integrable or not. The insets show that the variances in our calculations decay as a power law in $L$, with a power that is close to the expected $L^{-1}$ behavior. This shows that, with increasing system size, the overwhelming majority of eigenstates of local Hamiltonians are at the center of the spectrum (with a vanishing energy per site in our case). In what follows, we focus our scaling analyses on that region of the spectrum ($E_\alpha/L\simeq0$). 

To quantify the differences seen in Fig.~\ref{fig:eev_overlap} between the EEVs of observables in integrable and nonintegrable systems, we study the average of the absolute value of the eigenstate-to-eigenstate fluctuations~\cite{Kim2014Testing, 2DTFIM, Jansen2019Eigenstate} 
\beq \label{eq:eev_fluct}
\overline{|\delta O_{\alpha\alpha}|}=\overline{|O_{\alpha\alpha}-O_{\alpha+1\alpha+1}|} \, ,
\eeq
where the index $\alpha$ labels the eigenenergies $E_{\alpha}$ (sorted in increasing order), and $\overline{|\dots|}$ denotes an average over the central 20\% of eigenstates. To carry out an accurate comparison between our results for $\overline{|\delta O_{\alpha\alpha}|}$ and the ETH ansatz for quantum chaotic systems, a modification needs to be introduced to the ansatz in order to tailor it to our observables of interest. This modification is related to the fact that we focus on intensive operators that are defined via extensive sums, as seen in Eqs.~\eqref{eq:obs_A} and~\eqref{eq:obs_B}, in the presence of translational invariance. The Hilbert-Schmidt norm of those operators scales as $1/\sqrt{L}$~\cite{mierzejewski_vidmar_19}, as opposed to the $O(1)$ Hilbert-Schmidt norm one has in mind when writing Eq.~\eqref{eq:ETH}. As a result, for the diagonal part of our operators $\hat A$ and $\hat B$, Eq.~\eqref{eq:ETH} needs to be rewritten as:
\beq \label{eq:ETH_diag}
O_{\alpha\alpha} = O(E_{\alpha}) + \frac{e^{-S(E_{\alpha})/2}}{\sqrt{L}} f_{O} (E_{\alpha},0) R_{\alpha\alpha} \, .
\eeq
Since we are focusing on the regime $E_{\alpha}/L\simeq0$, in which $S(E_{\alpha})\simeq\ln(D)$ ($D$ is the dimension of the Hilbert space of the symmetry sector studied), we expect the average eigenstate-to-eigenstate fluctuations of $\hat{O}$ to be $\propto {(LD)}^{-1/2}$. For integrable systems, on the other hand, the average eigenstate-to-eigenstate fluctuations are expected to be proportional to $1/\sqrt{L}$~\cite{Alba}.

In Fig.~\ref{fig:fluct_trend}, we show the finite size scaling of $\overline{|\delta A_{\alpha\alpha}|}$ [Figs.~\ref{fig:fluct_trend}(a) and~\ref{fig:fluct_trend}(b)] and $\overline{|\delta B_{\alpha\alpha}|}$ [Figs.~\ref{fig:fluct_trend}(c) and~\ref{fig:fluct_trend}(d)] at two integrable [Figs.~\ref{fig:fluct_trend}(a) and~\ref{fig:fluct_trend}(c), $\lambda=0$] and two nonintegrable [Figs.~\ref{fig:fluct_trend}(b) and~\ref{fig:fluct_trend}(d), $\lambda=1$] points. For the nonintegrable points, we observe a near perfect scaling $\propto 1/\sqrt{LD}$ for both observables. This is in agreement with the ETH ansatz and indicates that the scaling is robust against the parameters of the model and the choice of observables. On the other hand, at integrability,  $\overline{|\delta A_{\alpha\alpha}|}$ and $\overline{|\delta B_{\alpha\alpha}|}$ exhibit a much slower decay with increasing system size, and also exhibit very strong finite-size effects. While we expect the decay to be $\propto 1/\sqrt{L}$~\cite{Alba}, this is not the exponent of the power law we find if we fit the data to $c_1 L^{-c_2}$. Instead, we have fitted the data to the function $\text{fit}(L)=c_1/\sqrt{L}+c_2/L$ and we find a reasonably good agreement. This suggests that higher powers of $1/\sqrt{L}$ still play an important role in the system sizes accessible to us through full exact diagonalization.

\begin{figure*}[!t]
\centering
\includegraphics[width=0.95\textwidth]{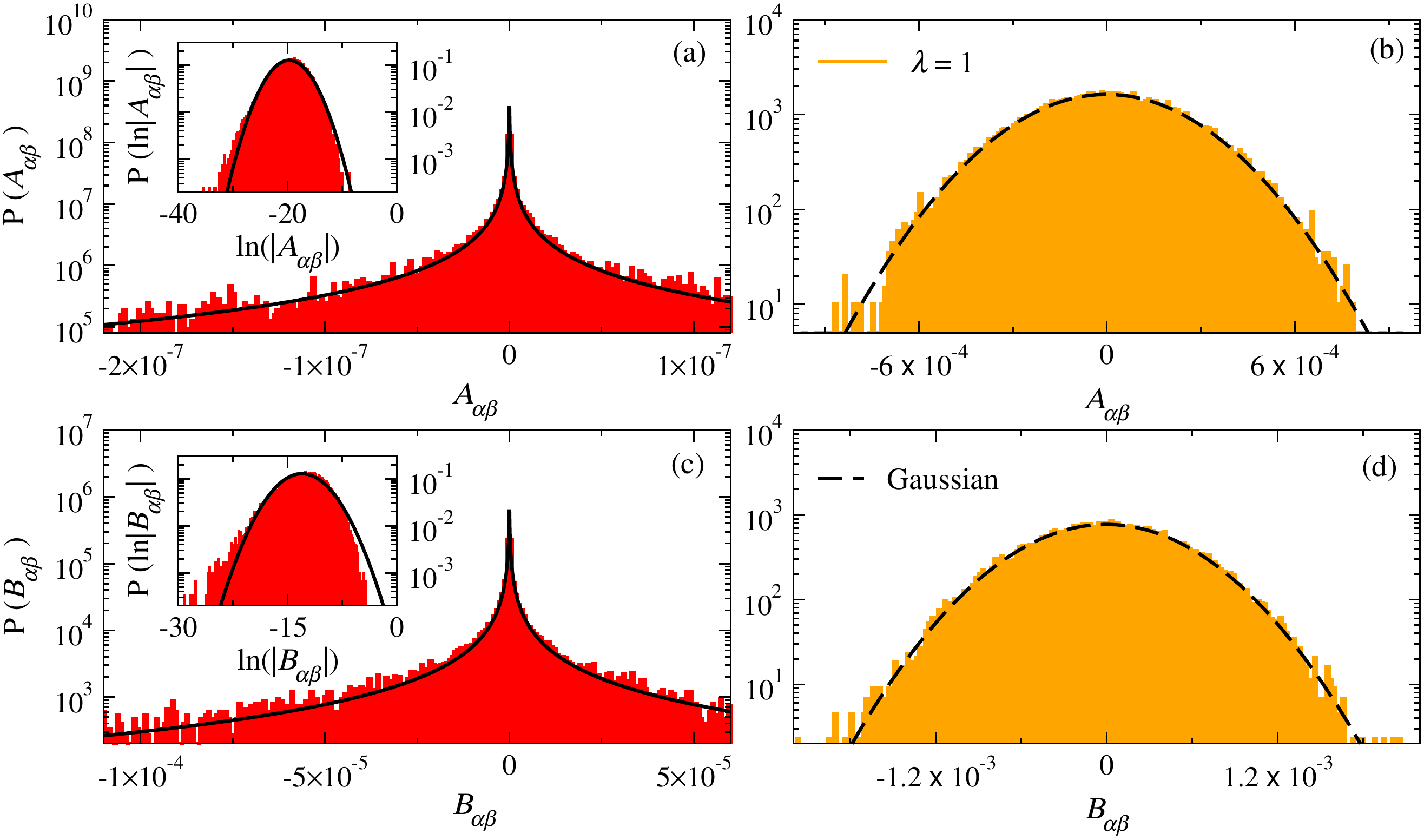}
\vspace{-0.1cm}
\caption{\label{fig:offdiag_distr} Probability distributions $P(O_{\alpha\beta})$ for observables $\hat{A}$ [(a) and (b)] and $\hat{B}$ [(c) and (d)] at integrable [(a) and (c), $\lambda=0$] and nonintegrable [(b) and (d), $\lambda=1$] points of Hamiltonian~\eqref{eq:xxz} with $\Delta=0.55$. We consider 200 energy eigenstates at the center of the spectrum. The insets in (a) and (c) show the probability distributions $P(\ln |O_{\alpha\beta}|)$, along with Gaussian distributions (continuous lines) with the same mean and variance. The continuous lines in the main panels in (a) and (c) are the corresponding log-normal distributions. The dashed lines in panels (b) and (d) are Gaussian distributions with the same mean and variance as the distributions $P(O_{\alpha\beta})$.} 
\end{figure*}

\section{Off-Diagonal Matrix Elements} \label{offdiag}

In this section, we study the off-diagonal matrix elements of observables in the eigenstates of Hamiltonian~\eqref{eq:xxz}. We focus on their distributions, scaling properties, and functional dependence of $|f_O(\bar{E}\simeq0,\omega)|^2$ on $\omega$, for eigenstates that are in the even-$Z_2$ even-$P$ sector within the $M^z=0$ and $k=0$ sector (see Sec.~\ref{model}).

\subsection{Distribution}

We first study the distribution of off-diagonal matrix elements of observables $\hat{A}$ and $\hat{B}$ within 200 energy eigenstates at the center of the spectrum of a chain with $L=26$ sites. In this eigenstate window, $\bar E=(E_{\alpha}+E_{\beta})/2\simeq0$ and $\omega=|E_{\alpha}-E_{\beta}|\simeq0$. For nonintegrable systems, this window is small enough to have $f_{O}(\bar{E},\omega)$ approximately constant, so that the probability distribution of $O_{\alpha\beta}$ is determined by $R_{\alpha\beta}$.

In Fig.~\ref{fig:offdiag_distr} we show the probability distributions of $A_{\alpha\beta}$ and $B_{\alpha\beta}$ at integrable and nonintegrable points of Hamiltonian~\eqref{eq:xxz}, with $\Delta=0.55$. At the nonintegrable point, Figs.~\ref{fig:offdiag_distr}(b) and~\ref{fig:offdiag_distr}(d) clearly show that the numerical results are well described by Gaussian distributions, as expected. At the integrable point, on the other hand, the distributions are fundamentally different [Figs.~\ref{fig:offdiag_distr}(a) and~\ref{fig:offdiag_distr}(c)]. While they also have approximately zero mean, they exhibit sharp peaks at the origin. Analyses of the distributions of $\ln|O_{\alpha\beta}|$ (shown in the insets) provide a better insight into the distributions of $O_{\alpha\beta}$. We find that, for our observables, the $\ln{|O_{\alpha\beta}|}$ distributions have skewed normal-like shapes [insets in Figs.~\ref{fig:offdiag_distr}(a) and~\ref{fig:offdiag_distr}(c)]. Gaussian distributions with the same mean and variance as our numerical results for $\ln|O_{\alpha\beta}|$, shown as continuous lines in the insets, illustrate the Gaussianity and skewness of the $\ln|O_{\alpha\beta}|$ distributions. The corresponding log-normal distributions, shown as continuous lines in the main panels, capture some of the features observed in the distributions of $O_{\alpha\beta}$ but fail to describe them quantitatively. Which distribution fully characterizes our results for $O_{\alpha\beta}$ at integrability remains a question for future studies.

We have also studied the distributions of $A_{\alpha\beta}$ and $B_{\alpha\beta}$ for $\bar{E}=(E_{\alpha}+E_{\beta})/2\simeq0$ and $\omega> 0$, obtaining qualitatively similar results to those shown in Fig.~\ref{fig:offdiag_distr} for $\bar{E}=(E_{\alpha}+E_{\beta})/2\simeq0$ and $\omega\simeq0$. Next, instead of reporting those distributions for $\omega> 0$, we study the ratio
\beq \label{eq:ratio} 
\Gamma_O(\omega)=\overline{|O_{\alpha\beta}|^2}/\overline{|O_{\alpha\beta}|}^2,
\eeq 
where $\alpha$ and $\beta$ are eigenstates that satisfy $|\bar{E}|/L\le0.025$, $\omega=|E_{\alpha}-E_{\beta}|$ takes values that vary throughout the entire spectrum, and $\overline{(\dots)}$ denotes a running average of the relevant quantity over eigenstates within a small $\omega$ window. If $O_{\alpha\beta}$ has a Gaussian distribution with zero mean, then $\Gamma_O(\omega)=\pi/2$ irrespectively of model parameters and the observable considered.

\begin{figure}[!t]
\centering \includegraphics[width=1.0\columnwidth]{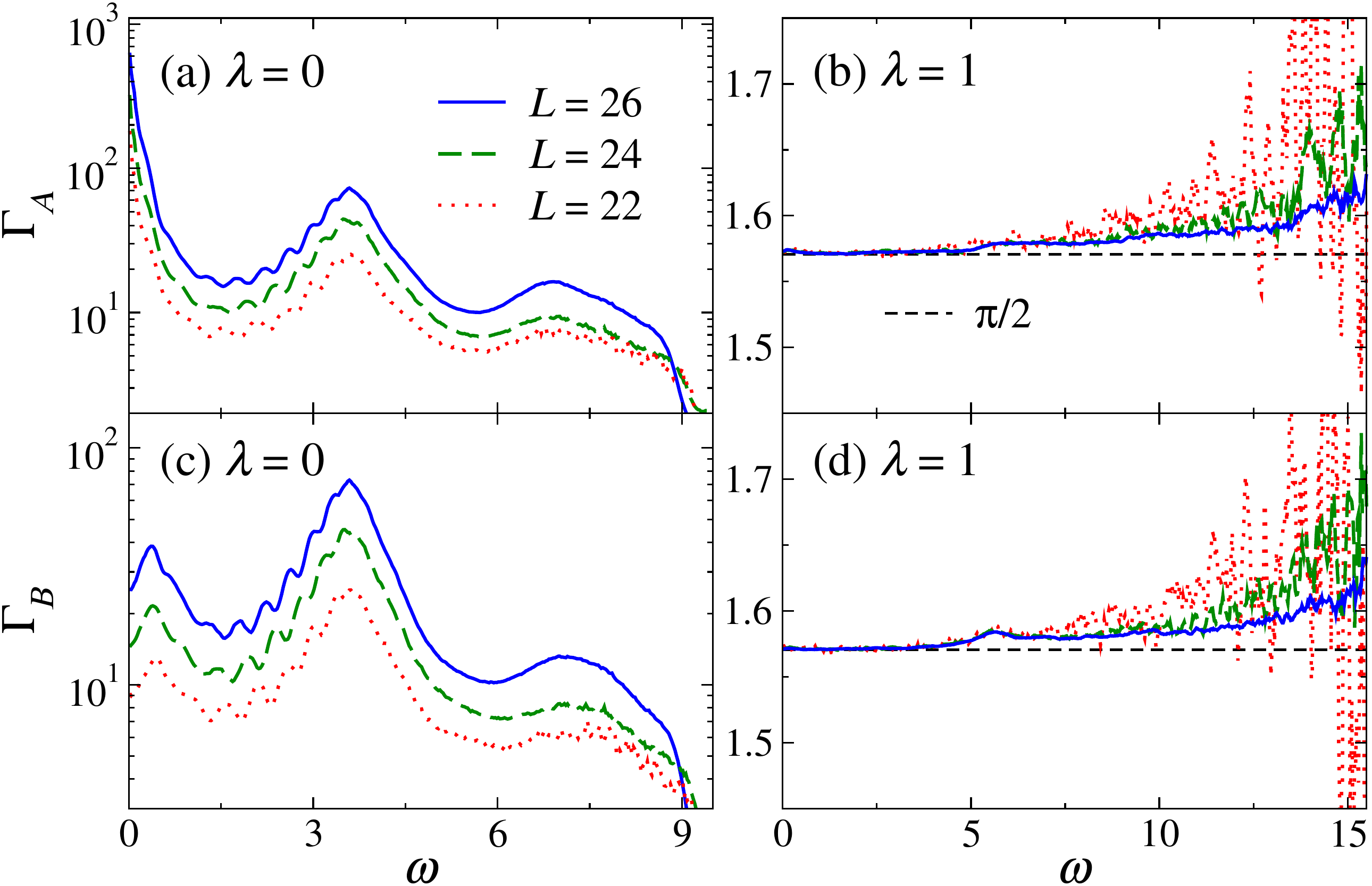}
\vspace{-0.4cm}
\caption{\label{fig:offdiag_ratio_omega} $\Gamma_O$ [see Eq.~\eqref{eq:ratio}] for observables $\hat{A}$ [(a) and (b)] and $\hat{B}$ [(c) and (d)] at integrable [(a) and (c), $\lambda=0$] and nonintegrable [(b) and (d), $\lambda=1$] points ($\Delta=0.55$) and different system sizes. We compute $\Gamma_O$ using eigenstates that satisfy $|\bar{E}|/L\leq 0.025$. The averages $\overline{|O_{\alpha\beta}|}$ and $\overline{|O_{\alpha\beta}|^2}$ are calculated with eigenstates in a window $\delta \omega = 0.175$ centered at points in $\omega$ separated by $\Delta\omega=0.025$.}
\end{figure}

At the nonintegrable point, Figs.~\ref{fig:offdiag_ratio_omega}(b) and~\ref{fig:offdiag_ratio_omega}(d) show that both $\Gamma_A(\omega)$ and $\Gamma_B(\omega)$ are almost indistinguishable from $\pi/2$ for $\omega\lesssim 5$. Discrepancies from $\pi/2$ can be seen for $\omega\gtrsim 5$. In this regime, we find in the next section that the variance of the off-diagonal matrix elements decreases rapidly with increasing $\omega$. For $5\lesssim\omega\lesssim 8$ in Figs.~\ref{fig:offdiag_ratio_omega}(b) and~\ref{fig:offdiag_ratio_omega}(d), $\Gamma_A(\omega)$ and $\Gamma_B(\omega)$ appear converged to results that could signal a small system-size-independent deviation from the Gaussian distribution prediction. However, finite-size effects are evident for $\omega\gtrsim 8$ [where $\Gamma_A(\omega)$ and $\Gamma_B(\omega)$ decrease with increasing system size] and could also be affecting the regime $5\lesssim\omega\lesssim 8$. Thus, whether $\Gamma_O(\omega)$ agrees with the Gaussian distribution prediction at high values of $\omega$ is something that requires future investigation. However, for $\omega\lesssim 5$, our results are a stringent test of the Gaussianity of the distributions of $O_{\alpha\beta}$ in the nonintegrable case.

In contrast, at the integrable point, Figs.~\ref{fig:offdiag_ratio_omega}(a) and~\ref{fig:offdiag_ratio_omega}(c) show that $\Gamma_A(\omega)$ and $\Gamma_B(\omega)$ depend on $\omega$, $L$, and the observable considered. This shows that the distribution of $O_{\alpha\beta}$ is not Gaussian at any $\omega$. A second point to be highlighted from the behavior of $\Gamma_O(\omega)$ at integrability is that, since $\Gamma_O(\omega)$ increases with increasing system size $L$, $\overline{|O_{\alpha\beta}|^2}$ decreases more slowly with increasing $L$ than $\overline{|O_{\alpha\beta}|}^2$. Since $\overline{|O_{\alpha\beta}|^2}$ is the quantity that enters in fluctuation dissipation relations~\cite{Khatami,ETH_review}, transport properties~\cite{Steinigeweg2013Eigenstate, Luitz2016Anomalous}, and heating rates under periodic driving~\cite{1907.04261}, in what follows we focus on the scaling of $\overline{|O_{\alpha\beta}|^2}$ with increasing system size, and on the smooth function that characterizes the dependence of $\overline{|O_{\alpha\beta}|^2}$ on $\omega$.  

\begin{figure}[!b]
\centering \includegraphics[width=1.0\columnwidth]{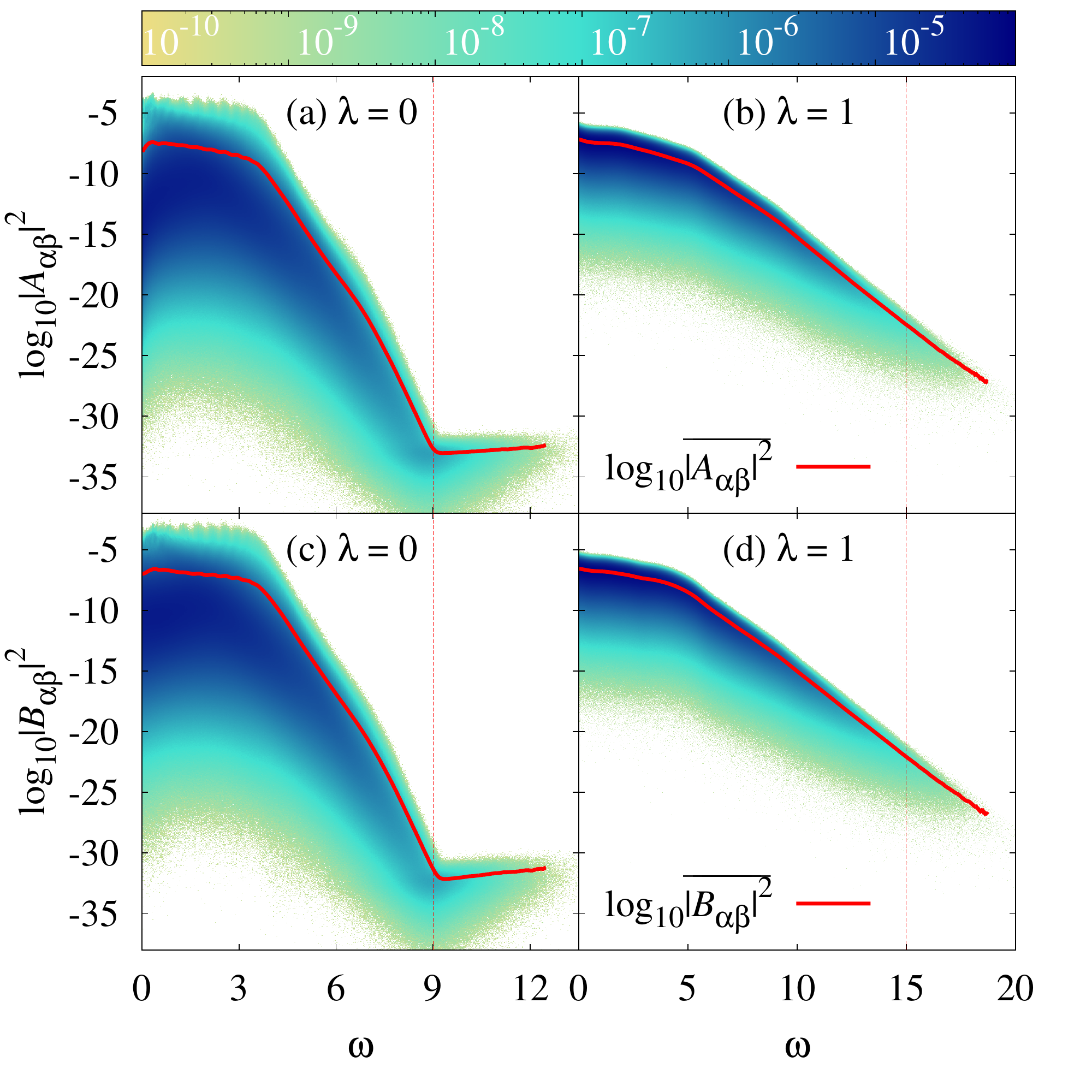}
\vspace{-0.6cm}
\caption{\label{fig:offdiag_omega} Normalized 2D histograms of $\log_{10}|O_{\alpha\beta}|^2$ vs $\omega$, for observables $\hat{A}$ [(a) and (b)] and $\hat{B}$ [(c) and (d)] at integrable [(a) and (c), $\lambda=0$] and nonintegrable [(b) and (d), $\lambda=1$] points of Hamiltonian~\eqref{eq:xxz} with $\Delta=0.55$. We consider eigenstates that satisfy $|\bar{E}|/L\le0.025$. The solid (red) lines are averages calculated using all the matrix elements in windows of widths $\delta\omega=0.175$ centered at points separated by $\Delta\omega=0.025$. The red dashed lines show the values of $\omega$ up to which results for $|O_{\alpha\beta}|^2$ are included in the scaling analyses of Fig.~\ref{fig:offdiagsqtrend}.} 
\end{figure}  

\subsection{Variance}

In Fig.~\ref{fig:offdiag_omega}, we show normalized distributions (color coded) of $\log_{10}|A_{\alpha\beta}|^2$ and $\log_{10}|B_{\alpha\beta}|^2$ versus $\omega$ at integrable and nonintegrable points of Hamiltonian~\eqref{eq:xxz}, for eigenstates that satisfy $|\bar{E}|/L\le0.025$. These results were obtained for $\Delta=0.55$ in chains with $L=26$. While for all values of $\omega$ the distributions are clearly different between the integrable and nonintegrable cases in that the former have a much broader support, neither of them has an increasing fraction of matrix elements that vanish with increasing system size as in quadratic models~\cite{Khatami}. This means that one can define a meaningful average $\overline{|O_{\alpha\beta}|^2}$ for each value of $\omega$, which, given that $\overline{O_{\alpha\beta}}\simeq0$, is also the variance of $O_{\alpha\beta}$. In Fig.~\ref{fig:offdiag_omega}, we also plot the variances of $O_{\alpha\beta}$. Again, while they are quantitatively different between the integrable and nonintegrable cases, the overall behavior is qualitatively similar. They exhibit a slow decay at intermediate values of $\omega$ ($0.5\lesssim\omega\lesssim4$) and a fast decay at larger values of $\omega$.

\begin{figure}[!t]
\centering \includegraphics[width=1.0\columnwidth]{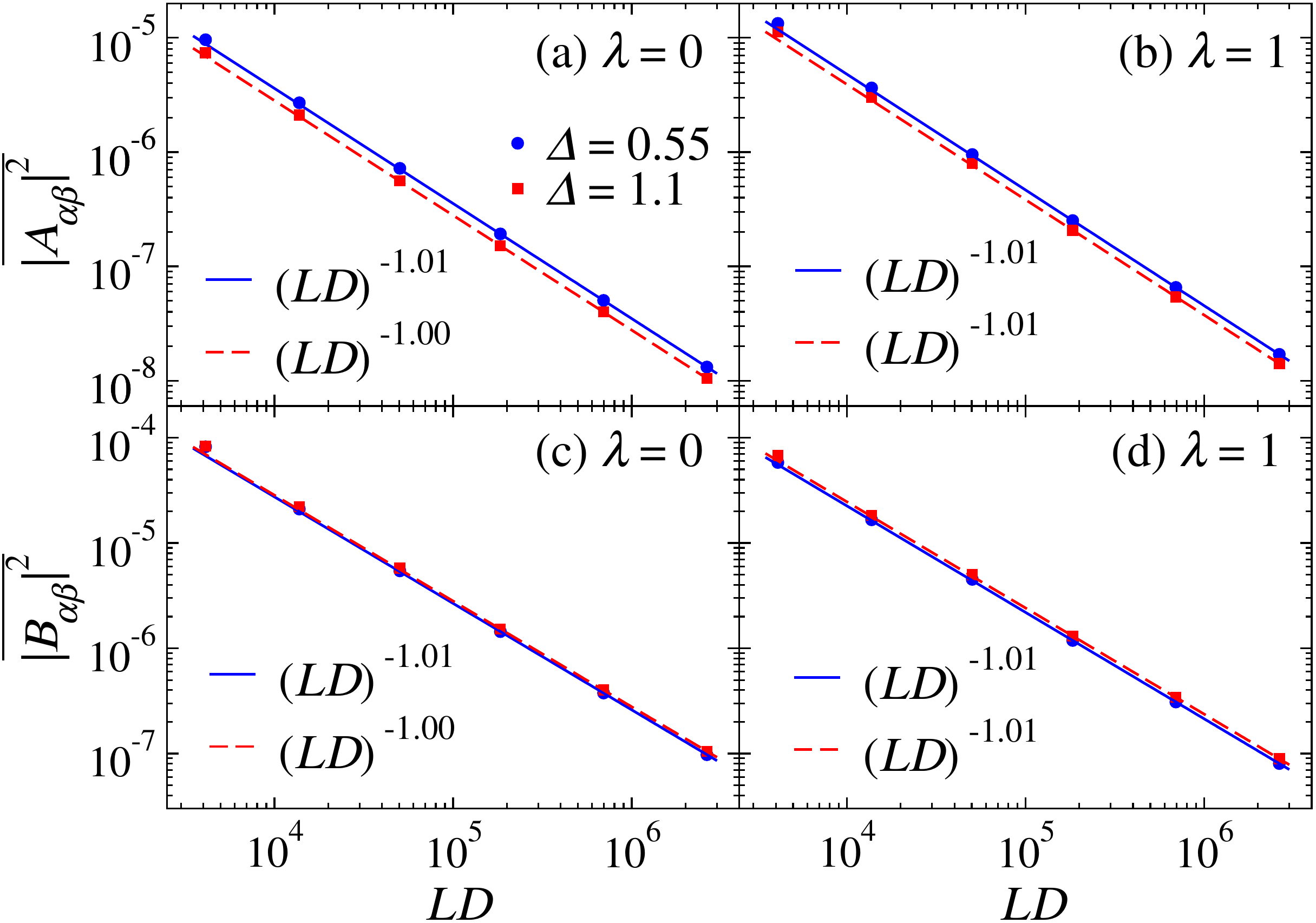}
\vspace{-0.3cm}
\caption{\label{fig:offdiagsqtrend} Scaling of $\overline{|A_{\alpha\beta}|^2}$ [(a) and (b)] and $\overline{|B_{\alpha\beta}|^2}$ [(c) and (d)] vs $LD$ at integrable [(a) and (c), $\lambda=0$] and nonintegrable [(b) and (d), $\lambda=1$] points ($\Delta=0.55$ and 1.1) of Hamiltonian~\eqref{eq:xxz}. The straight lines show power-law fits to the results for $L=22$ through $L=26$. The average over $|O_{\alpha\beta}|^2$ for different system sizes was calculated using eigenstates that satisfy $|\bar{E}|/L\leq 0.025$. For the integrable cases we used eigenstates with $\omega<9$ [see the vertical dashed lines in Figs.~\ref{fig:offdiag_omega}(a) and~\ref{fig:offdiag_omega}(c)], while for the nonintegrable ones we used eigenstates with $\omega<15$ [see the vertical dashed lines in Figs.~\ref{fig:offdiag_omega}(b) and~\ref{fig:offdiag_omega}(d)]. These ranges of $\omega$ were the ones populated with matrix elements for the system sizes considered.} 
\end{figure}

Next, we study how $\overline{|O_{\alpha\beta}|^2}$ scales with increasing the system size (and, hence, with increasing the dimension $D$ of the Hilbert space). In the quantum chaotic case, we expect the scaling to be the one prescribed by the ETH. However, as we did when studying the fluctuations of the diagonal matrix elements in Sec.~\ref{diag}, we need to update the ETH ansatz to account for the fact that our translationally invariant operators $\hat{A}$ and $\hat{B}$ have a Hilbert-Schmidt norm that scales as $1/\sqrt{L}$. The ETH ansatz for the off-diagonal matrix elements of $\hat{A}$ and $\hat{B}$ has the form
\beq \label{eq:ETH_offdiag} 
O_{\alpha\beta}=\frac{e^{-S(\bar{E})/2}}{\sqrt{L}} f_O(\bar{E},\omega)R_{\alpha\beta},
\eeq
where $\bar{E}\equiv (E_{\alpha}+E_{\beta})/2$, $\omega=E_{\alpha}-E_{\beta}$, and $S(\bar E)$ is the thermodynamic entropy at energy $\bar E$. Since we are focusing on the regime $\bar{E}/L\simeq0$, in which $S(\bar{E})\simeq\ln(D)$, we expect $\overline{|O_{\alpha\beta}|^2}\propto (LD)^{-1}$, where $D$ is the dimension of the symmetry sector studied. Figures~\ref{fig:offdiagsqtrend}(b) and~\ref{fig:offdiagsqtrend}(d) show that this is indeed the way $\overline{|O_{\alpha\beta}|^2}$ scales with increasing system size. More remarkably, as shown in Figs.~\ref{fig:offdiagsqtrend}(a) and~\ref{fig:offdiagsqtrend}(c), the same is equally true for the integrable case as conjectured in Ref.~\cite{1907.04261}.

\begin{figure}[!t]
\centering \includegraphics[width=1.0\columnwidth]{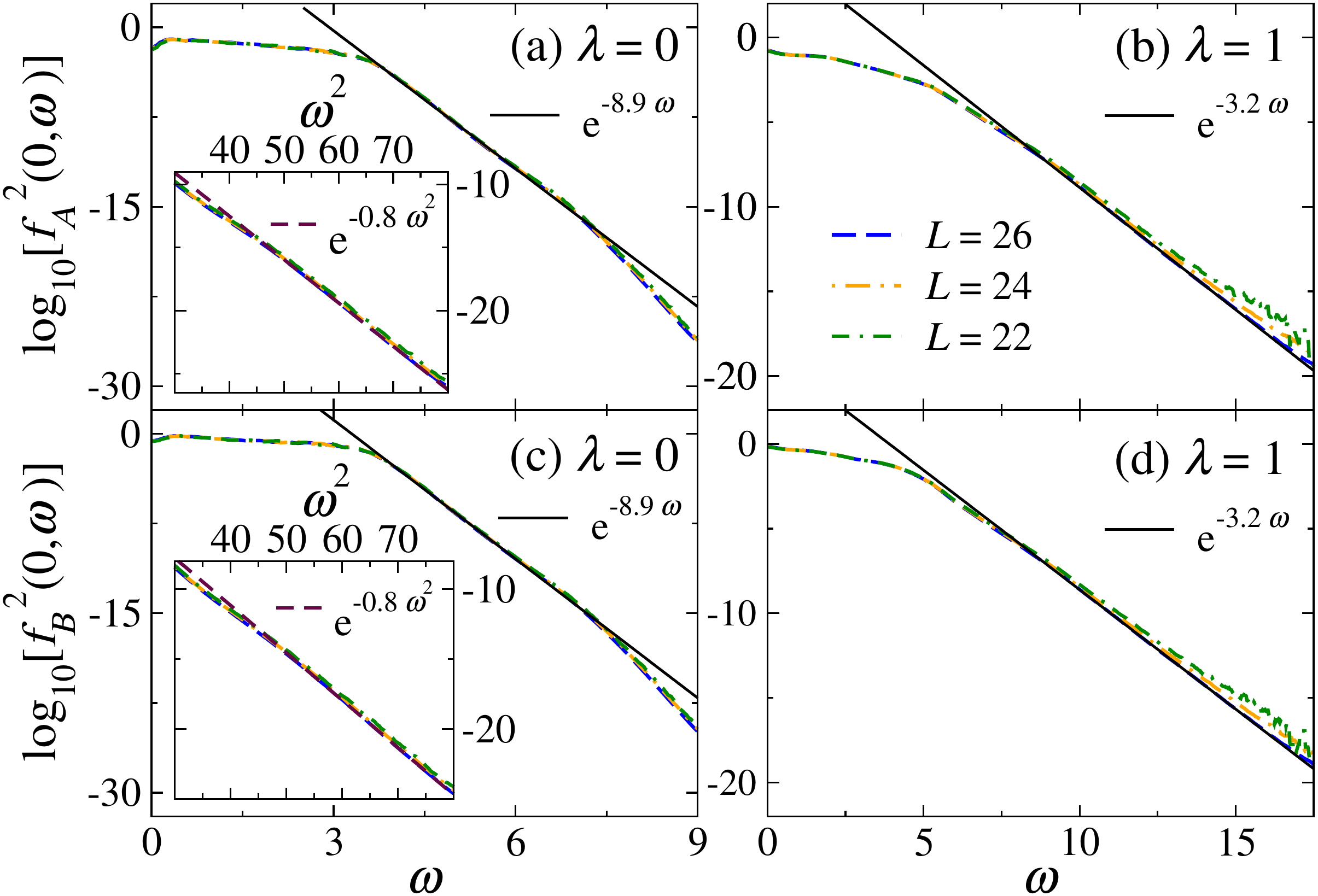}
\vspace{-0.4cm}
\caption{Smooth functions $|f_O(\bar{E}\simeq0,\omega)|^2$ [see Eq (\ref{eq:f2})] for observables $A$ [(a) and (b)] and $B$ [(c) and (d)] vs $\omega$ at integrable [(a) and (c), $\lambda=0$] and nonintegrable [(b) and (d), $\lambda=1$] points ($\Delta=0.55$) for different system sizes $L$. The straight continuous lines are exponential fits $\propto \exp(-a\omega)$ to $|f_O(\bar{E}\simeq0,\omega)|^2$ for $L=26$. The insets in (a) and (c) show $|f_O(\bar{E}\simeq0,\omega)|^2$ vs $\omega^2$, at high $\omega$, for different system sizes $L$. The straight dashed lines are Gaussian fits $\propto \exp(-b\omega^2)$ to $|f_O(\bar{E}\simeq0,\omega)|^2$ for $L=26$.} \label{fig:offdiag_omega_scaled} 
\end{figure}

Armed with this knowledge, we can now extract the smooth function $|f_O(\bar{E}\simeq0,\omega)|^2$, which is independent of system size for nonvanishing values of $\omega$~\cite{ETH_review}, that characterizes $\overline{|O_{\alpha\beta}|^2}$. That function, calculated as   
\beq \label{eq:f2} 
|f_O(\bar{E}\simeq0,\omega)|^2 = LD\,\overline{|O_{\alpha\beta}|^2} \; ,
\eeq
is plotted in Fig.~\ref{fig:offdiag_omega_scaled} for our two observables of interest, at integrable [Figs.~\ref{fig:offdiag_omega_scaled}(a) and~\ref{fig:offdiag_omega_scaled}(c)] and nonintegrable [Figs.~\ref{fig:offdiag_omega_scaled}(b) and~\ref{fig:offdiag_omega_scaled}(d)] points, for the three largest system sizes studied. As advanced, we obtain nearly perfect data collapse for different system sizes. At the nonintegrable point, $|f_O(\bar{E}\simeq0,\omega)|^2$ exhibits finite-size effects for $\omega\gtrsim 10$. This is the result of running out of spectrum in our finite-system calculations. However, Figs.~\ref{fig:offdiag_omega_scaled}(b) and~\ref{fig:offdiag_omega_scaled}(d) show that, with increasing system size, the results for different values of $L$ agree over a larger range of values of $\omega$. Overall, the results in Fig.~\ref{fig:offdiag_omega_scaled} strongly suggest that the function $|f_O(\bar{E},\omega)|^2$ is a well-defined smooth function of $\bar{E}$ and $\omega$ (independent of system size for nonvanishing values of $\omega$) both in interacting integrable and nonintegrable systems.

Finally, we would like to discuss the decay of $|f_O(\bar{E}\simeq0,\omega)|^2$ for large values of $\omega$, after the slow decay mentioned before. [The behavior of $|f_O(\bar{E}\simeq0,\omega)|^2$ for very small values of $\omega$ is of much interest, but it is also more challenging to address computationally~\cite{ETH_review}. It will not be discussed here.] As shown in Fig.~\ref{fig:offdiag_omega_scaled}, at both the integrable and nonintegrable points, $|f_A(\bar{E}\simeq0,\omega)|^2$ and $|f_B(\bar{E}\simeq0,\omega)|^2$ can be well described by exponentials after the initial slow decay. This is well known for nonintegrable systems~\cite{ETH_review}, and we find that it also occurs for integrable ones. However, at the integrable point we find yet another regime beyond the exponential one. As shown in the insets in Figs.~\ref{fig:offdiag_omega_scaled}(a) and~\ref{fig:offdiag_omega_scaled}(c), we find that the final decay of $|f_A(\bar{E}\simeq0,\omega)|^2$ and $|f_B(\bar{E}\simeq0,\omega)|^2$ to zero (within machine precision) is nearly perfectly Gaussian. Whether a Gaussian decay occurs at nonintegrable points for larger values of $\omega$ than those accessible to us via full exact diagonalization remains an open question. Upon replotting the $|f_O(\bar{E}\simeq0,\omega)|^2$ data reported in Ref.~\cite{Jansen2019Eigenstate} for the Holstein polaron model, which was shown to be a quantum chaotic model, we found that it can be well described by a Gaussian decay (without any exponential part). This suggests that Gaussian decays of $|f_O(\bar{E},\omega)|^2$ with $\omega$ are not unique to integrable models.
  
\section{Summary} \label{conc}

We studied the bipartite von Neumann entanglement entropy and matrix elements of local operators in highly excited eigenstates of interacting integrable (the spin-1/2 XXZ chain) and nonintegrable models. 

For the average entanglement entropy over all eigenstates in the zero magnetization sector, we found that the leading term is extensive at interacting integrable points with a coefficient of the volume-law that is smaller (for nonvanishing ratios $L_{\rm A}/L$) than the universal $\ln 2$ coefficient in quantum chaotic models. Finite-size scaling analyses suggested that the coefficient at $L_{\rm A}/L=1/2$, and for arbitrary ratios $L_{\rm A}/L$ (not reported), is (almost) independent of the XXZ chain anisotropy parameter $\Delta$, and that it is very close or equal to that of translationally invariant free fermionic Hamiltonians. Since the average entanglement entropy over all eigenstates is dominated by eigenstates at ``infinite temperature'', the presence, or lack thereof, of interactions (and their values) at integrability may play no role in the leading extensive term. What may be essential is that the system is integrable so that it has an underlying quasiparticle description. Hence, we find it plausible that all translationally invariant (two-state per site) integrable models (noninteracting and interacting) have the same average entanglement entropy for any given ratio $L_{\rm A}/L$. If this is the case, then one could think of two universality classes for the average entanglement entropy of all ``q-bit'' based physical Hamiltonians, (translationally invariant) free fermions characterizing integrable models, and random matrices characterizing nonintegrable ones. 

For the diagonal matrix elements of observables at the center of the spectrum and at interacting integrable points, we showed evidence that the support does not vanish with increasing system size and that the average eigenstate-to-eigenstate fluctuation vanishes as a power law in system size. At nonintegrable points, however, both the support and the average eigenstate-to-eigenstate fluctuation vanish exponentially with increasing system size. For the off-diagonal matrix elements with $\bar E=(E_{\alpha}+E_{\beta})/2$ at the center of the spectrum, we showed that at interacting integrable points they follow a distribution that is close to (but not quite) log-normal, and that their variance is a well-defined function of $\omega=E_{\alpha}-E_{\beta}$ whose magnitude scales as $1/D$, where $D$ is the Hilbert space dimension. The latter is a known property of the off-diagonal matrix elements of observables in nonintegrable models, which, however, exhibit a Gaussian distribution. We also studied the smooth function $|f_O(\bar{E}\simeq0,\omega)|^2$ that characterizes the variance and contrasted its behavior at interacting integrable and nonintegrable points. It was recently argued that this function can be measured in experiments with periodically driven systems, both nonintegrable and interacting integrable ones, by studying how heating rates change when changing the frequency of the drive~\cite{1907.04261}. An interesting open question is whether the Bethe ansatz can be used to analytically learn about the smooth function $|f_O(\bar{E},\omega)|^2$ in interacting integrable systems. This would pave the way for an analytic understanding of the effect of interactions in the matrix elements of observables in many-body quantum systems. 

\acknowledgements
We acknowledge discussions with S. Gopalakrishnan and M. Mierzejewski. This work was supported by the National Science Foundation under Grant No.~PHY-1707482 (T.L., K.M., and M.R.), and the Slovenian Research Agency (ARRS), Research core fundings Grants No.~P1-0044 and No.~J1-1696 (L.V.).

\bibliographystyle{biblev1}
\bibliography{refs,refs_ee}

\end{document}